\documentclass[preprint,nofootinbib,floats,aps]{revtex4}
\usepackage{graphicx}
\usepackage{epsfig}
\def\be{\begin{equation}}
\def\ee{\end{equation}}
\def\br{\begin{eqnarray}}
\def\er{\end{eqnarray}}
\def\brn{\begin{eqnarray*}}
\def\ern{\end{eqnarray*}}

\def\R {{{\cal R}}}
\def\ie{{\em i.e., }}
\def\nn{\nonumber }
\def\rf#1{{(\ref{#1})}}

\def\e{\epsilon}

\def\s {{\sigma}}

\def\w {{\omega}}
\def\go{\rightarrow}

\def\x{\times}

\def\etal{{\it et al. }}
\begin{document}

\title{A Reanalysis of the LSND Neutrino Oscillation Experiment}

\author{A. Samana$^{1,2,3}$, F. Krmpoti\'c$^{2,3,4}$,
A. Mariano$^{3,4}$ and  R. Zukanovich Funchal$^{2}$}
\affiliation{
$^1$ Centro Brasileiro de Pesquisas F\'{\i}sicas,
CEP 22290-180, Rio de Janeiro,Brazil}
\affiliation{$^2$ Instituto de F{\'\i}sica,  Universidade de S{\~a}o Paulo
C.P. 66.318, 05315-970, S{\~a}o Paulo, Brazil}
\affiliation{$^3$Departamento de F\'{\i}sica, Facultad de Ciencias Exactas,
Universidad Nacional de La Plata, La Plata, CP 1900, Argentina}
\affiliation{$^4$Instituto de F\'{\i}sica La Plata-CONICET,
115 y 49, La Plata, CP 1900, Argentina.
}

\begin{abstract}
We reanalyse the LSND neutrino oscillation results in the framework of
the Projected Quasiparticle Random Phase Approximation (PQRPA), which
is the only RPA model that treats  the Pauli Principle correctly, and
accounts satisfactorily for great majority of the weak decay
observables  around $^{12}C$.
We have found that the employment of the PQRPA inclusive DIF
$^{12}C(\nu_e,e^-)^{12}N$ cross-section, instead of the CRPA used by
the LSND collaboration in the ${\nu}_\mu \go{\nu}_e$ oscillations
study of the $1993-1995$ data sample, leads to the following:
1) the oscillation probability is increased from
$(0.26\pm 0.10 \pm 0.05)\%$ to $(0.33\pm 0.10 \pm 0.13)\%$, and
2) the previously found consistence between the
$(\sin^2 2\theta, \Delta m^2)$ confidence level regions for the
${\nu}_\mu\go {\nu}_e$ and the $\bar{\nu}_\mu\go \bar{\nu}_e$
oscillations is significantly diminished.  These effects are not
due to the difference in the uncertainty ranges for the
neutrino-nucleus cross-section, but to the difference in the
cross-sections themselves.

\end{abstract}
 \maketitle

Several recent
experiments~\cite{Ath96,Ath98,Agu01,Fuk98,Aha05,Ara04,Ahn03}
strongly suggest that neutrinos oscillate.
This means that a neutrino of a certain flavor
(e.g.  ${\nu}_\mu$) transforms as it propagates into a neutrino of
another flavor (e.g. ${\nu}_e$), violating the conservation of the
lepton number.
For this to happen, the simplest and most widely
accepted explanation is that neutrinos have masses and mixing.  There
are evidences of transitions for three different $\Delta m^2$: $\sim
8\times 10^{-5}$ eV$^2$ (solar), $\sim 2.5\times 10^{-3}$ eV$^2$
(atmospheric) and $\sim 1 $ eV$^2$ (LNSD), which cannot all be
understood in the context of three neutrino oscillations.  Normally,
the {\it Liquid Scintillator Neutrino Detector} (LSND) results are not
taken into account when fitting neutrino
oscillation data.  Nevertheless, one has to try to understand the
real significance of the LSND measurements, specially because
$\Delta m^2 \sim $ 1 eV$^2$ is of particular interest to astrophysics
and cosmology.
The  LSND experiment took place over six calendar years
finding evidence for the appearance of electron-antineutrinos $\bar{\nu}_e$
at the 3.3$\sigma$ level~\cite{Ath96,Agu01},
and  at lesser significance they have observed as well
hints for the appearance of electron-neutrinos $\nu_e $~\cite{Ath98,Agu01}.
 The first of these signals is the main
LSND result and the second weaker signal was used as a consistency check.
The positive results in both channels were
interpreted in a two-flavor framework as  transitions
between  the weak eigenstates $\nu_\mu$ ($\bar \nu_\mu$) and $\nu_e$
($\bar \nu_e$) driven by masses and mixing.
In fact, quantum mechanics dictates that in this case
the  normally observed weak eigenstates   $(\nu_\mu, \nu_e)$
 can oscillate between each other with probability
\be
P_{\nu_\mu\go \nu_e}=
\sin^2(2\theta)
\sin^2\left(1.27 \; \Delta m^2 \frac{L_\nu}{E_\nu} \right),
\label{1}\ee
 if they are composed of a mixture of mass eigenstates $(\nu_1, \nu_2)$.  Here
$\theta$ is the mixing angle between the mass and flavor bases,
$\Delta m^2=m_1^2-m_2^2$ is the $\nu_1$ and $\nu_2$ mass squared
differences in eV$^2$, $L_\nu$ is the baseline, the distance in meters
travelled by the neutrino from the source to the detector, and $E_\nu$
is the neutrino energy in MeV.

The combination of the LSND data with other compelling evidences for
 neutrino oscillations, stemming from atmospheric~\cite{Fuk98},
 solar~\cite{Aha05}, KamLAND~\cite{Ara04} reactor, and
 K2K~\cite{Ahn03} accelerator neutrino experiments, cannot be
 adequately explained in the standard three-neutrino picture with
 $CPT$ conservation, and this issue is considered to be a big
 challenge for neutrino phenomenology~\cite{Sch03,Mal04,Fuk94}.
Models with four light neutrinos (the extra neutrino being
 sterile)~\cite{sterile} or $CPT$ violation~\cite{cptv} with three
 neutrinos have been proposed to accommodate all neutrino data. However,
 in both cases, recent analyses show that neither scenario
 provides a satisfactory description of the data~\cite{not-four, not-cptv}.

In the LSND experiment the neutrinos $\nu_\mu$ come from the decay of
$\pi^+$ in flight ({\it decay in flight}, DIF), whereas the neutrinos
$\nu_e$ and the antineutrinos ${\bar \nu}_\mu$ come from the decay of
$\mu^+$ at rest ({\it decay at rest}, DAR), \ie \brn
&&\pi^+\go\mu^++\nu_\mu\hspace{0.9cm}\pi^+\go\mu^++\nu_\mu \\
&&\hspace{4.5cm}\downarrow \\ &&\hspace{4.5cm}e^+ +\nu_e+{\bar
\nu}_\mu.  \\ &&\hspace{1cm}{\rm DIF}\hspace{3cm}{\rm DAR} \ern

The search for the DAR $\bar{\nu}_\mu \go \bar{\nu}_e$
oscillations~\cite{Ath96,Agu01} involves the measurement of the
reaction $p(\bar{\nu}_e,e^+)n$, which has a large and well known
cross section.  The events are identified by  detecting both the
$e^+$ and the $2.2$ MeV $\gamma$-ray from the reaction $p(n,\gamma)d$.
On the other hand, the signature for the DIF $\nu_\mu \go
\nu_e$ oscillations~\cite{Ath98,Agu01} is marked by the presence
of an isolated high energy electron ($60<E_e^{\rm DIF}<200$ MeV)
 in the detector.
It is produced by the charge-exchange reaction $^{12}C(\nu_e,e^-)^{12}N$,
which takes place throughout the tank,
the cross section of which $\sigma_{e}$ is, as yet, not well established.
 The lower and upper energy
cuts for $E^{\rm DIF}_e$ were chosen in such a way as to be above the
Michel electron end point of $52.8$ MeV and below the point where the
beam-off background starts to increase rapidly and the signal becomes
negligible.

There are two LSND studies of the  DIF $\nu_\mu \go  \nu_e$ oscillations.
The first analysis was done on the  $1993-1995$ data sample~\cite{Ath98},
which gave a total of $N^{osc}_{\nu_e}=18.1\pm 6.6\pm 4.0$ oscillation events,
corresponding to a transition probability
\be
P_{\nu_\mu\go \nu_e}^{exp}=(2.6\pm 1.0 \pm 0.5)\x10^{-3},
\label{2}\ee
 when the cross-section $\sigma_{e}$ predicted by Kolbe \etal within
the Continuum Random Phase Approximation (CRPA) is used~\cite{Kol94}.
In the second search, the $1996-1998$ data sample~\cite{Agu01} was
included as well, with reduced DIF flux and higher beam-off background
compared to the $1993-1995$ data.
The reason for this modification  lies in the fact that in this study
first priority  was given to the DAR $\bar{\nu}_\mu
\go \bar{\nu}_e$ oscillations, which have been analysed jointly.
 Moreover,
for $\sigma_e$ has been employed in this occasion two different models. Namely,
 the   shell model (SM)
 estimate,  done by  Hayes and Towner ~\cite{Hay00}, for the DAR region,
and  a relativistic Fermi gas model for the DIF region.
The resulting total excess was $N^{osc}_{\nu_e}=8.1\pm 12.2 \pm 1.7$  events,
 yielding
\be
 P_{\nu_\mu\go \nu_e}^{exp}=(1.0\pm 1.6 \pm 0.4)\x10^{-3}.
\label{3}\ee
The aim of the present work is to explore the role played by
these nuclear structure effects in the  delimitation  of the neutrino
 parameters for the DIF $\nu_\mu \go  \nu_e$
oscillations.\footnote{Accurate knowledge of the
$\nu$ cross-section, and the related
observables, plays an important role for the next generation of experiments.
Various target nuclei, like C, O, Fe, Ar, Pb, $\cdots$, are normally
(and presumably will be) employed to provide the detector mass.}

Before proceeding, and to make more clear the objective
of the present work,  it is convenient to discuss briefly
 the flux-averaged  exclusive cross sections
\br
\overline{\s}_\ell^{\rm exc}&=& \int_0^{E_{\nu_\ell}^{max}} dE_{\nu}
\s_\ell(E_\ell=E_\nu-\Delta,1^{+}_1){\Phi}_\ell(E_{\nu})
\label{4}\er
and the inclusive cross sections
\br
\overline{\s}_\ell^{\rm inc}&=& \int_0^{E_{\nu_\ell}^{max}} dE_{\nu}
\s_\ell(E_{\nu}) {\Phi}_\ell(E_{\nu}),
\label{5}\er
where
\be
\s_\ell(E_{\nu})=\sum_{J^{\pi}_f}
\s_\ell(E_\ell=E_\nu-\w_{J_{f}^\pi},J_{f}^\pi);~ ~ \ell=e,\mu.
 \label{6}\ee
The spin and parity  dependent cross section $\s_\ell(E_\ell,J_{f}^\pi)$
is given by \cite[(2.19)]{Krm05}, $\w_{J_{f}}$ are the excitation
energies in $^{12}$N relative to the ground state
in $^{12}$C, and $\Delta\equiv\w_{1^{+}_1}=17.3$~MeV.  The energy
integration  for electrons is carried out in the
DAR  interval $m_e+\w_{J_{f}}\le \Delta^{\rm DAR}_{J_f}\le
E_{\nu_e}^{max}= 52.8 $ MeV, and  for muons in the
DIF  interval up to $m_\mu+\w_{J_{f}}\le \Delta^{\rm DIF}_{J_f}\le
E_{\nu_\mu}^{max}=300 $ MeV.
\footnote{In order to invert the summation on $J_{f}^\pi$ and the
integration on  $dE_{\nu}$,   we have extended the lower limit of
integration in \rf{4} from $m_\ell+\w_{J_{f}}$ to zero  by defining
$\s_\ell(E_\ell=E_\nu-\w_{J_{f}},J_{f}^\pi)\equiv 0$ for
$E_\ell<m_\ell$.}
$\Phi_\ell(E_{\nu})$ is the normalized  neutrino flux; for $\nu_e$
it is approximated by the Michel spectrum, and
for $\nu_\mu$ that from Ref. \cite{LSND} was used.

\begin{table}[h]
\caption{ Calculated and experimental flux-averaged  exclusive
$\overline{\s}_{e, \mu}^{\rm exc} $,
and inclusive $\overline{\s}_{\mu}^{\rm inc }$
cross section for the
$^{12}C(\nu_e,e^-)^{12}N$ DAR reaction (in units of $10^{-42}$~cm$^{2}$)
and for the $^{12}C(\nu_\mu,\mu^-)^{12}N$ DIF reaction (in units of
$10^{-40}$ cm$^{2}$).
The CRPA calculations~\cite{Kol94} were used in the first LSND analysis
 on the 1993-1995 data sample~\cite{Ath98}, and the  SM
calculations from Ref.~\cite{Hay00} in the second LSND oscillation
search \cite{Agu01}.  The  listed  PQRPA results correspond to the
calculations performed  with the relativistic corrections included
\cite{Krm05}. One alternative SM result as well as the RPA and
 QRPA results from Ref.~\cite{Vol00} are also shown.}
\label{table1}
\newcommand{\cc}[1]{\multicolumn{1}{c}{#1}}
\renewcommand{\tabcolsep}{ 0.6pc} 
\renewcommand{\arraystretch}{1.2} 
\bigskip
\begin{tabular}{l|cccc}\hline
&$\overline {\s}_e^{\rm exc}$
&$\overline {\s}_e^{\rm inc }$
&$\overline {\s}_\mu^{\rm exc}$
&$\overline {\s}_\mu^{\rm inc}$
\\\hline
\underline{Theory}&&&&\\
CRPA \cite{Kol94} &36.0, 38.4&42.3, 44.3&2.48, 3.11 &21.1, 22.8\\
SM \cite{Hay00}   &7.9 &12.0&0.56&13.8\\
PQRPA \cite{Krm05}&8.1&18.6&0.59&13.0\\
SM \cite{Vol00}   &8.4&16.4&0.70 &21.1\\
RPA  \cite{Vol00} &49.5&55.1&2.09&19.2\\
QRPA \cite{Vol00} &42.9&52.0&1.97&20.3\\
\hline
\underline{Experiment}&&&&\\
Ref. \cite{Ath97}&$9.1 \pm 0.4\pm 0.9 $&$14.8 \pm0.7\pm 1.4$&&\\
Ref. \cite{Ath97a}&&&$0.66\pm 0.1\pm 0.1$&$12.4 \pm 0.3 \pm 1.8$\\
Ref. \cite{Aue01}&$8.9 \pm 0.3\pm 0.9 $&$13.2 \pm0.4\pm 0.6 $&&\\
Ref. \cite{Aue02a}&&&$0.56\pm 0.08\pm 0.10$&$10.6\pm 0.3 \pm 1.8 $\\
\hline
\end{tabular}
\label{tabla1}\end{table}

The experimental data for the exclusive and inclusive
 cross sections, given in  Table \ref{table1}, show that the   DAR and DIF
 processes  are of quite different nature: while the first
 one is dominated in proportion of $2/3$ by the Gamow-Teller (GT)
 transition to  the ground state $1^+_1$ in $^{12}N$, the second one
populates  almost entirely  the excited states  through
the forbidden transitions.
It is quite a difficult task for the
nuclear structure models to describe both cross sections simultaneously.

The SM treats correctly the Pauli Principle within the $p$-shell,
which is crucial for the correct distribution of the GT strength,
whereas the predictions for high-lying states are less certain because
of the truncation of the model space.
 In fact, the SM calculation performed by Hayes and
 Towner~\cite{Hay00}  reproduces fairly well several data.
 But,  in a later SM study,  Volpe~\etal\cite{Vol00} noted that
this concordance could  be an artifact because the employed model space
 was not large enough to exhaust the charge-exchange
sum rules. More, the same authors have shown that
when   a more extended space is  employed the SM cross
sections are increased exceeding the
experimental LSND result.

The RPA like models include high-lying one-particle one-hole
excitations, but very frequently completely fail to account for
the amount and distribution of the GT strength as can be seen from
Table \ref{table1}. This is the reason why
the CRPA is unable to explain  the weak processes
($\beta$-decays, $\mu$-capture, and neutrino  induced reactions)
among the ground states of the triad $\{{{^{12}B},{^{12}C},{^{12}N}}\}$:
a rescaling factor of the order of $4$ is needed
to bring the calculations and the data to agree~\cite{Kol94}, and  a
subsequent {\it ad hoc} inclusion of partial occupancy of the
$p_{1/2}$ subshell reduces this factor to less than $2$ \cite{Aue97,Kol99}.
It is still more relevant here that the CRPA overestimates
 the inclusive  $^{12}C(\nu_\mu,\mu^-)^{12}N$ cross-section
 with $\nu_\mu$ coming from the DIF of $\pi^+$   by about
 $50\%$~\cite{Ath97a} or more~\cite{Aue02a}, because
one can assume
that the DIF $^{12}C(\nu_e,e^-)^{12}N$ cross section,
which gauges  the $\nu_\mu \go  \nu_e$ oscillations, is
 affected in the same proportion. This assumption comes
from  the universality of the weak interaction and
was  done in the first LSND analysis~\cite{Ath98}.
\footnote{Since the work of O'Connell, Donelly and
  Walecka~\cite{Con72}
we know that electron and muon cross sections
 differ for low neutrino energy, but tend to
merge for high neutrino energy.}

 Thus, it might be interesting to reanalyse the LSND results
 in the framework of  the Projected
Quasiparticle Random Phase Approximation (PQRPA) \cite{Krm02},  which
is the only RPA model that treats correctly the Pauli Principle,
explaining in this way the distribution of the GT strength.
To achieve  this it was  imperative both: a) to include the BCS
correlations,  and b) to perform the particle number projection.
Under these conditions most of  the weak decay
observables around $^{12}C$ are within $20\%$ of the PQRPA
predictions. This happens, for instance,
 with: 1) the B(GT)-values to $^{12}N$  and $^{12}B$, 2)  the exclusive
muon captures  to the
 $1_1^+$,  $2_1^+$, $1_1^-$ and $2_1^-$ states, as well as the
inclusive muon capture   in  $^{12}B$, and 3)
the exclusive cross sections $\overline{\s}_e^{\rm exc}$
and  $\overline{\s}_\mu^{\rm exc}$ and the inclusive cross section
$\overline{\s}_\mu^{\rm exc}$~\cite{Krm02,Krm05}.  The only exception is the
 inclusive cross section, $\overline{\s}_e^{\rm inc *}=\overline{\s}_e^{\rm inc}
-\overline{\s}_e^{\rm exc}$, for which the  PQRPA value,  10.5
(in units of $10^{-40}$ cm$^2$),
is more than 100\% larger that the experiment result,
 $ 4.3\pm 0.4\pm 0.6$~\cite{Aue01}. From the nuclear structure point
 of view the theoretical evaluation of this quantity is a peculiarly
delicate and subtle issue and therefore  deserves a special comment.
In fact from  Table VI in Ref.\cite{Krm05} it can be seen that in the
PQRPA case  $\overline{\s}_e^{\rm inc *}$ is  build up from the  interplay
of GT strength not contained in the $1^+_1$ state,
the Fermi (F) transitions to
the $0^+$ states, and the first forbidden transitions  to the  $1^-$ and
 $2^-$ states.  All these quantities are relatively small and
 evaluating them precisely is a very difficult task.
 Then one should not be
surprised by the most recent  SM calculation~\cite{Vol00} which
yields a result ($\overline{\s}_e^{\rm inc *}=8.3$) which is twice
as large as
that obtained in the previous SM study:
$\overline{\s}_e^{\rm inc *}=4.1$~\cite{Hay00}.
The CRPA result $\overline{\s}_e^{\rm inc *}=6.3$~\cite{Kol94},
very likely does not contain any GT and F strengths as it should,
and therefore, in this case,  the agreement with the experiment could be
accidental.

\begin{figure}[h]
\begin{center}
   \leavevmode
   \epsfxsize = 9cm
     \epsfysize = 9cm
 \epsffile{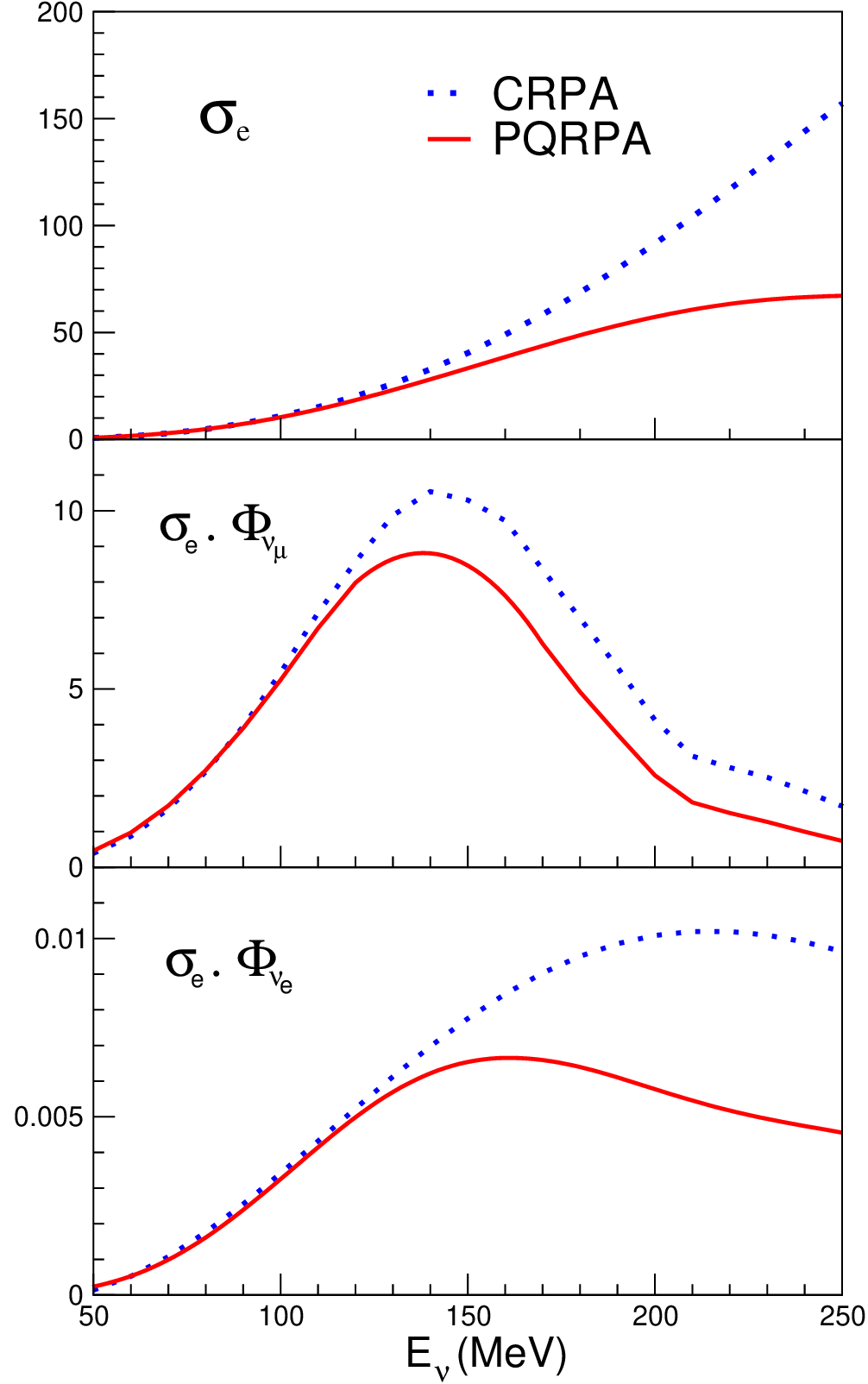}
\end{center}
\caption{
(Color online) Comparison
between  the  CRPA and PQRPA results for:  $\sigma_e (E_\nu)$ in units
of $10^{-40}$ cm$^2$ (upper panel), and,  in units of  $10^{-52}$ POT
$^{-1}$ MeV $^{-1}$, for
$\sigma_e(E_\nu)\Phi_{\nu_\mu}$ (middle panel) and
$\sigma_e(E_\nu)\Phi_{\nu_e}$ (lower panel).}
\label{figure1}\end{figure}

We will limit our attention only to the $1993-1995$ data
sample~\cite{Ath98}, which, as mentioned before, yields a more
defined signal for the oscillation events.
 The  experimental oscillation probability can be written as
\br
P_{{\nu_\mu}\go{\nu_e}}^{exp}&=&\frac{N_{\nu}}
{{\e f_n }
\langle \sigma \Phi_{{\nu_\mu}}\rangle}
-\frac{\langle\sigma \Phi_{{\nu_e}}\rangle}{\langle \sigma
\Phi_{{\nu_\mu}}\rangle},
\label{7}\er
where  the $\nu_e$ flux  (from now on) is defined as
\be
\Phi_{\nu_e}=\Phi_{\nu_e}^{\mu^+}+\Phi_{\nu_e}^{\pi^+},
\label{8}\ee
with the fluxes $\Phi_{\nu_e}^{\mu^+}$ and $\Phi_{\nu_e}^{\pi^+}$ coming,
respectively, from the DIF decays  $\pi^+\go e^++\nu_e$ and
$\mu^+\go e^++\nu_e+{\bar \nu}_\mu$.
$f_n=(9.23 \x 10^{22})~ \cdot (5.4 \x 10^{30})$, with the first
 quantity  being the number of protons on target (POT),
while the second one
is the fiducial volume (number of molecules of $CH_2$ in the detector tank).
$N_{\nu}=N_{\nu}^{osc}+N_{\nu}^{bg}=27.7\pm 6.9$
is the total number of beam-excess events measured by LSND, and $\e$ is
the event selection efficiency. The averaged inclusive cross-sections are
\br
\langle\sigma \Phi_{\nu_\ell}\rangle
&=&\sum_{J^\pi_f}\int_{E^<_{J^\pi_f}}^{E^>_{J^\pi_f}}
\s_e(E_e=E_\nu-\w_{J_f},J_f^\pi)\Phi_{\nu_\ell} dE_\nu;~ ~ \ell=e,\mu,
\label{9}\er
where
$E^<_{J^\pi_f}=60~{\rm MeV}+\w_{J_f^\pi}$ and
$E^>_{J^\pi_f}=200~{\rm MeV}+\w_{J_f^\pi}$.
In order to simplify the numerical calculations which follow,
instead of using the exact equations \rf{8},  we will employ here the
approximate ones:
\br
\langle\sigma \Phi_{\nu_\ell}\rangle
&=&\int_{E^<}^{E^>}
\sigma_e(E_\nu)\Phi_{\nu_\ell} dE_\nu ;~ ~ \ell=e,\mu,
\label{10}\er
where $\sigma_e(E_\nu)$ is given by \rf{5}, and
$E^<=60~{\rm MeV}+\Delta$, and $E^>=200~{\rm MeV}+\Delta$.
We have  verified numerically that the  equations \rf{10}  reproduce the
equations \rf{9} up to a few per cent.
 The neutrino fluxes $\Phi_{\nu_\mu},\Phi_{\nu_e}^{\pi^+}$ and
$\Phi_{\nu_e}^{\mu^+}$ were adopted from the Ref.~\cite{Ath98}.
The  CRPA
and PQRPA results for  $\sigma_e (E_\nu)$, $\sigma_e(E_\nu)\Phi_{\nu_\mu}$ and
$\sigma_e(E_\nu)\Phi_{\nu_e}$ are confronted in
 Fig. \ref{figure1}, as a function of $E_\nu$.

\bigskip

The systematic error associated with the PQRPA cross-section is taken
to be $20 \%$, based on  our theoretical uncertainties (see
Tables V, VI and VII in~\cite{Krm05}),
 and  agreement between measured data and theoretical
predictions for the weak decay observables involving the $^{12}$C nucleus
\cite{Krm02,Krm05}.
Therefore, considering the same uncertainties
 as in the LSND search \cite{Ath98} in the selection of
 $\epsilon$ ($12 \%$) and in the flux $\Phi_{\nu_\mu}$
($15 \%$), we end up  with a total
systematic error of $28\%$, which yields
$N^{osc}_\nu=21.5 \pm 6.6 \pm 8.5$.
In this way the PQRPA result for the oscillation probability
turns out to be:
\be
P_{\nu_\mu\go \nu_e}^{exp}=(3.3\pm 1.0 \pm 1.3)\x10^{-3}.
\label{11}\ee
The difference  when compared to the CRPA result
\rf{2} is due to the difference
in the electron
cross-section, as evidenced in Fig. \ref{figure1}.
\begin{figure}[t]
\begin{center}
\vspace{1.cm}
   \leavevmode
   \epsfxsize = 7cm
     \epsfysize = 6.cm
    \epsffile{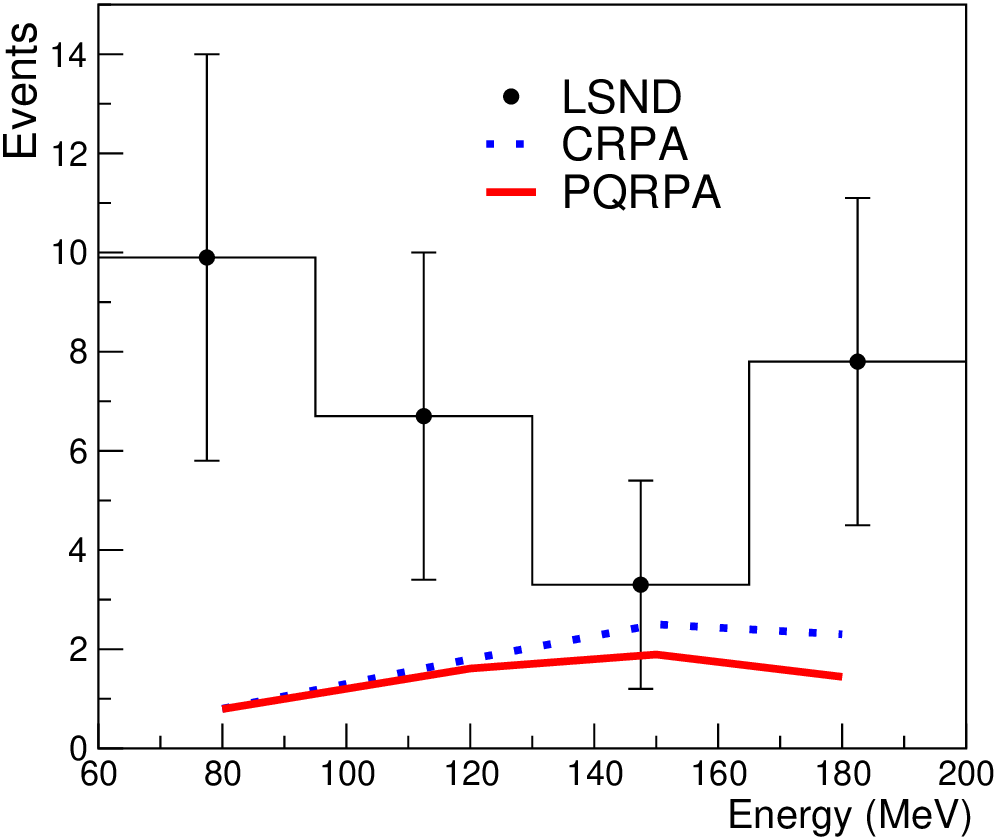}
\end{center}
\caption{(Color online) The energy distribution (in 4 energy bins) of
the LNSD excess events, $N_\nu(i)$, together with the corresponding
experimental errors  $\delta N_\nu(i)$ (vertical lines) and the energy
intervals   $E_\nu(i)$ (horizontal lines). The theoretical
CRPA and PQRPA values
for the expected background events, ${\tilde N}_\nu^{bg}(i)$, are
shown as well.}
\label{figure2}\end{figure}

In order to determine a confidence region
in the  $(\sin^2 2\theta, \Delta m^2)$ parameter space we proceed in
the same manner as in  Ref.\cite{Ath98}. First
we  rearrange  the data for  energy distribution
of the excess events (see \cite[Fig. 29]{Ath98})
 in four equal energy bins $N_\nu(i)$, as shown in Fig. \ref{figure2}.
 Next, we minimize the $\chi^2$ function
\be
\chi^2=\sum_{i=1}^4\left[
\frac{N_\nu(i)-{\tilde N}_\nu(i)}
{\delta N_\nu(i)}\right]^2,
\label{12}\ee
where ${\tilde N}_\nu(i)= {\tilde N}^{osc}_\nu(i)+{\tilde N}^{bg}_\nu(i)$,
with
\br
 {\tilde N}^{osc}_\nu(i)
&=& \epsilon f_n \int_{E_\nu(i)} \sigma(E_\nu)
\R(E_\nu) \Phi_{\nu_\mu} (E_\nu)P_{\nu_\mu\go \nu_e}
 dE_\nu,
\nn\\
 {\tilde N}^{bg}_\nu(i) &=&
\epsilon f_n \int_{E_{\nu}(i)} \sigma(E_\nu)
\R(E_\nu) \Phi^{bg}_{\nu_e}(E_\nu) dE_\nu,
\label{13}\er
where $P_{\nu_\mu\go \nu_e}$ is a function of $E_\nu$,
$\sin^2(2\theta)$ and $\Delta m^2$, and is defined in \rf{1}.
We include the resolution function \cite{Bah97},
\br
\R(E_\nu)&=&
\frac{1}
{ \sqrt{2 \pi}\varepsilon(E_\nu) }\int_{E_<}^{E_>}
\exp\left[-{1\over2}
\left({E'_\nu -E_\nu\over\varepsilon(E_\nu)}\right)^2\right]
dE'_\nu,
\er
which  takes into account the
probability for  finding the electron inside the window of detection,
with $\varepsilon(E_\nu)=0.06E_\nu$ being the experimental energy
resolution \cite{Ath98}.
\begin{figure}[t]
\begin{center}
   \leavevmode
   \epsfxsize = 9cm
     \epsfysize = 12cm
    \epsffile{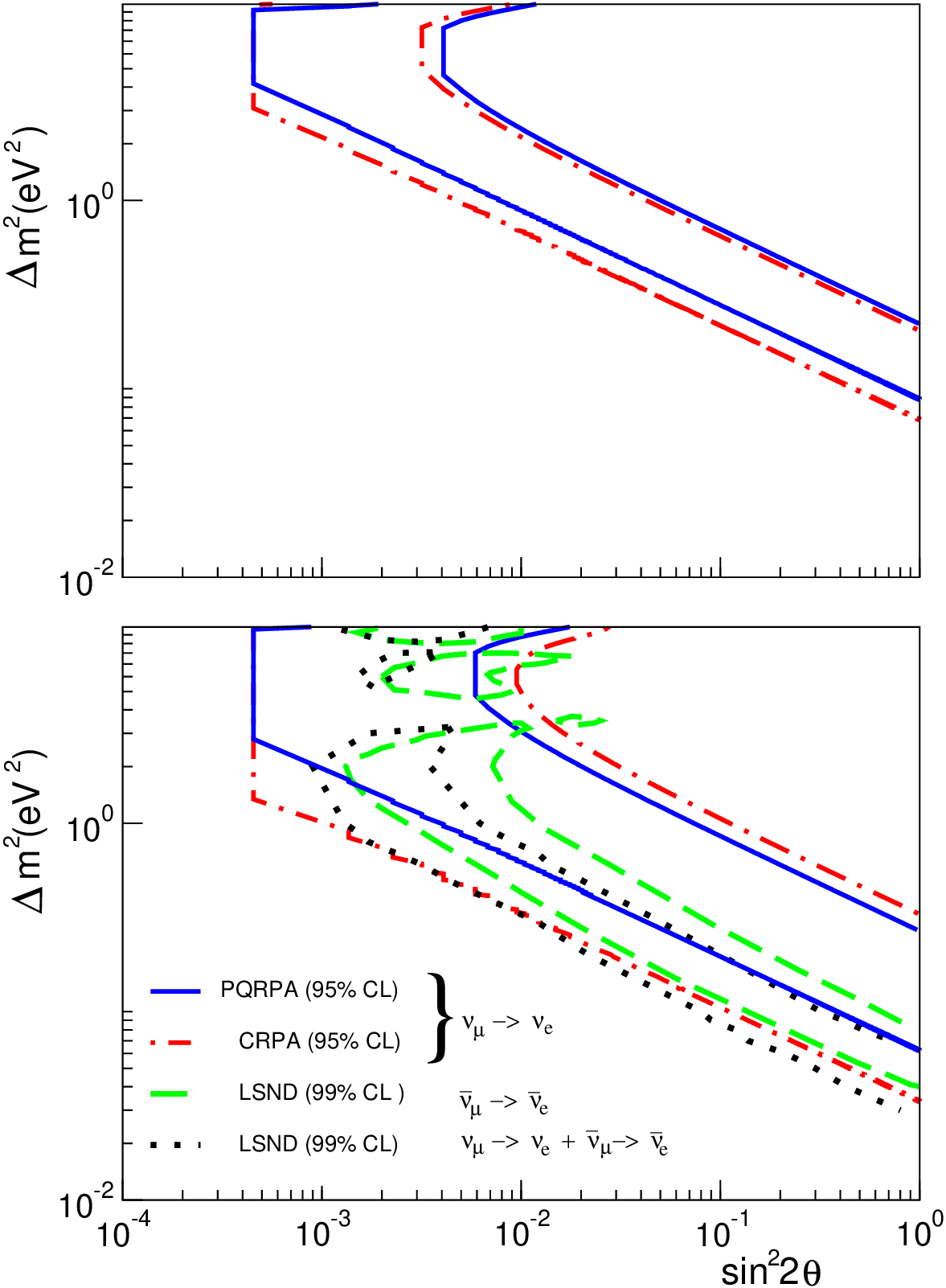}
\end{center}
\caption{(Color online)
Regions in the  neutrino oscillation parameter space.
 In the upper panel the results for $\nu_\mu\go \nu_e$ oscillations without the
inclusion of the systematic uncertainty are shown, while the lower
panel shows those with
the uncertainty included,
as described in the text.}
\label{figure3}\end{figure}

To set the confidence levels (CL) we  used the raster scan method
\cite{Fel98}: for each value of $\Delta m^2$, a best fit is found for
$\sin^2 2\theta$. At each $\Delta m^2$, $\chi^2$ is calculated as a
function of $\sin^2 2\theta$. The 1D confidence interval in
$\sin^2 2\theta$ at $\Delta m^2$ is composed of all points having
a $\chi^2$ within 3.84 of the minimum value (3.84 is the two-sided 95
$\%$ CL for a distribution $\chi^2$ with one degree of freedom). The
confidence region in the $(\sin^2 2\theta,\Delta m^2)$ is the union
of all these intervals.

Our  $(\sin^2 2\theta, \Delta m^2)$ oscillation parameter fits
for the DIF channel ${\nu}_\mu \go {\nu}_e$, corresponding to both
the CRPA~\cite{Kol94}
and PQRPA~\cite{Krm05}
cross-sections,  are  shown in Fig. \ref{figure3}, along with the favored
regions for the LSND DAR measurement for $\bar{\nu}_\mu \go
\bar{\nu}_e$ \cite{Ath96}.
In order to better understand the consequences of using
different cross-sections, the confidence
regions obtained with (lower panel) and without (upper panel)
inclusion of the systematic uncertainties, are displayed separately.
In the calculation with the CRPA cross-section these uncertainties
are taken to be the same as in the LSND search \cite{Ath98}, \ie
of $22 \%$ for the positive side, which shifts the parameter space
downwards,  and of
$45\%$ for the negative side, which shifts the parameter space
upwards. On the other hand,
an  uncertainty of $28 \%$ is used for both negative and positive
side,  when the PQRPA cross-section is employed.

We see that, when the  systematic uncertainties are considered, the
CRPA $95\%$
CL region  fully comprises  the $99\%$
CL region for the $\bar{\nu}_\mu \go \bar{\nu}_e$ oscillations,
which is in essence the result obtained by the  LSND collaboration
 \cite{Ath98}, Contrarily, this does not happens in the PQRPA case
where the overlapping between the two regions is only marginal.
It is important to stress that  the  ${\nu}_\mu \go{\nu}_e$  region
is dragged towards the  $\bar{\nu}_\mu \go \bar{\nu}_e$ region by the
positive side uncertainty, while the role played by the negative side
uncertainty is of minor importance.
For the sake of completeness
 the result
of the joint ${\nu}_\mu \go{\nu}_e$
and  $\bar{\nu}_\mu \go \bar{\nu}_e$  oscillation parameter fit
over
 $(\sin^2 2\theta, \Delta m^2)$ plane for the complete
 $1993-1998$ data sample~\cite{Agu01}, is also displayed
in Fig. \ref{figure3}.

In summary,  we have found that the employment of
  a smaller inclusive DIF $^{12}C(\nu_e,e^-)^{12}N$ cross-section,
than  the one used  by the  LSND collaboration~\cite{Ath98}
in the
 ${\nu}_\mu \go{\nu}_e$  oscillations study
of the 1993-1995 data sample, leads to the following consequences:
1)  the oscillation probability $P_{\nu_\mu\go \nu_e}^{exp}$
is increased, and
2) the previously found consistence between the $(\sin^2 2\theta,
 \Delta m^2)$ confidence level regions for the ${\nu}_\mu\go {\nu}_e$
 and the $\bar{\nu}_\mu\go \bar{\nu}_e$ oscillations is diminished.
 More, these effects are not due to the difference in the uncertainty
 ranges for the neutrino-nucleus cross-section, but to the difference
 in the cross-sections themselves, and are quite significant when the
 PQRPA is used instead of the CRPA.  Thus,  precise
 knowledge of the nuclear structure involved in the $\nu$-nucleus
 cross-section, could play an important role in the delimitation of
 the neutrino parameters for the DIF $\nu_\mu \go \nu_e$ oscillations.

{\bf Acknowledgements}

F.K. and A.M. are fellows of the CONICET, Argentina. A.S.
acknowledges support received from Conselho Nacional de Ci\^encia e
Tecnologia (CNPq) and Funda\c{c}\~ao de Amparo \`a Pesquisa do
Estado do Rio de Janeiro (FAPERJ).
R.Z.F. thanks CNPq and Funda\c{c}\~ao de Amparo \`a Pesquisa do
Estado de S\~ao Paulo (FAPESP) for partial financial support.
We are extremely grateful to Gordana
Tadi\'c for her very careful reading of the manuscript.


\end{document}